\documentclass[12pt]{article}
\usepackage{graphicx}
 
\def \bea{\begin{eqnarray}}
\def \beq{\begin{equation}}
\def \eea{\end{eqnarray}}
\def \eeq{\end{equation}}
 
\def \ok{\overline{K}^0}

\begin{document} 
\rightline{EFI 18-10} 
\rightline{TECHNION-PH-2018-6}
\rightline{arXiv:1808.03720} 
\rightline{September 2018} 
\bigskip
\centerline{Addendum to ``Overview of $\Lambda_c$ decays''}
\bigskip
 
\centerline{Michael Gronau\footnote{gronau@physics.technion.ac.il}} 
\centerline{\it Physics Department, Technion -- Israel Institute of Technology} 
\centerline{\it 32000 Haifa, Israel} 
\medskip 
 
\centerline{Jonathan L. Rosner\footnote{rosner@hep.uchicago.edu}} 
\centerline{\it Enrico Fermi Institute and Department of Physics} 
\centerline{\it University of Chicago, 5640 S. Ellis Avenue, Chicago, IL 60637} 
\medskip

\centerline{Charles G. Wohl\footnote{cgwohl@lbl.gov}}
\centerline{\it Physics Division, Lawrence Berkeley Laboratory}
\centerline{\it 1 Cyclotron Road, Berkeley, CA 94720}

\begin{quote}
An earlier analysis of observed and anticipated $\Lambda_c$ decays \cite{GR18}
is provided with a table of inputs and a figure denoting branching fractions.
This addendum is based on the 2018 compilation in Ref.\ \cite{PDG18}
and employs a statistical isospin model to estimate branching fractions for
as-yet-unseen decay modes.
\end{quote} 

The decays of the charmed baryon $\Lambda_c$ \cite{PDG18} appear to be within
about 10\% of fully mapped out \cite{GR18} when a statistical isospin model
\cite{stat,Peshkin:1976kw} is used to estimate branching fractions for
as-yet-unseen decay modes.  In this addendum to Ref.\ \cite{GR18}
we display graphically the
modes which have been seen and those anticipated.  Part of the $\sim 10\%$
shortfall may be composed of decay modes such as $\Lambda_c \to \Lambda^*
\ell^+ \nu_\ell$, where $\Lambda^*$ is an excited resonance, or may be due to
a shortcoming in the statistical isospin model.  Cabibbo-suppressed modes
appear to be less well-represented by known or anticipated decays, and are
worthy of more experimental study.  In order for this analysis to serve as a
model-independent counterpart to a Particle Data Group analysis of $D_s$
decays \cite{PDGDs}, measurements of inclusive branching fractions of
$\Lambda_c$ decays need to be undertaken. [An example is the result from
BESIII \cite{Ablikim:2018jfs}, ${\cal B}(\Lambda_c \to \Lambda + X) =
(38.2^{+2.8}_{-2.2} \pm 0.8)\%$.]
\medskip 

\begin{figure}
\begin{center}
\includegraphics[width=0.47\textwidth]{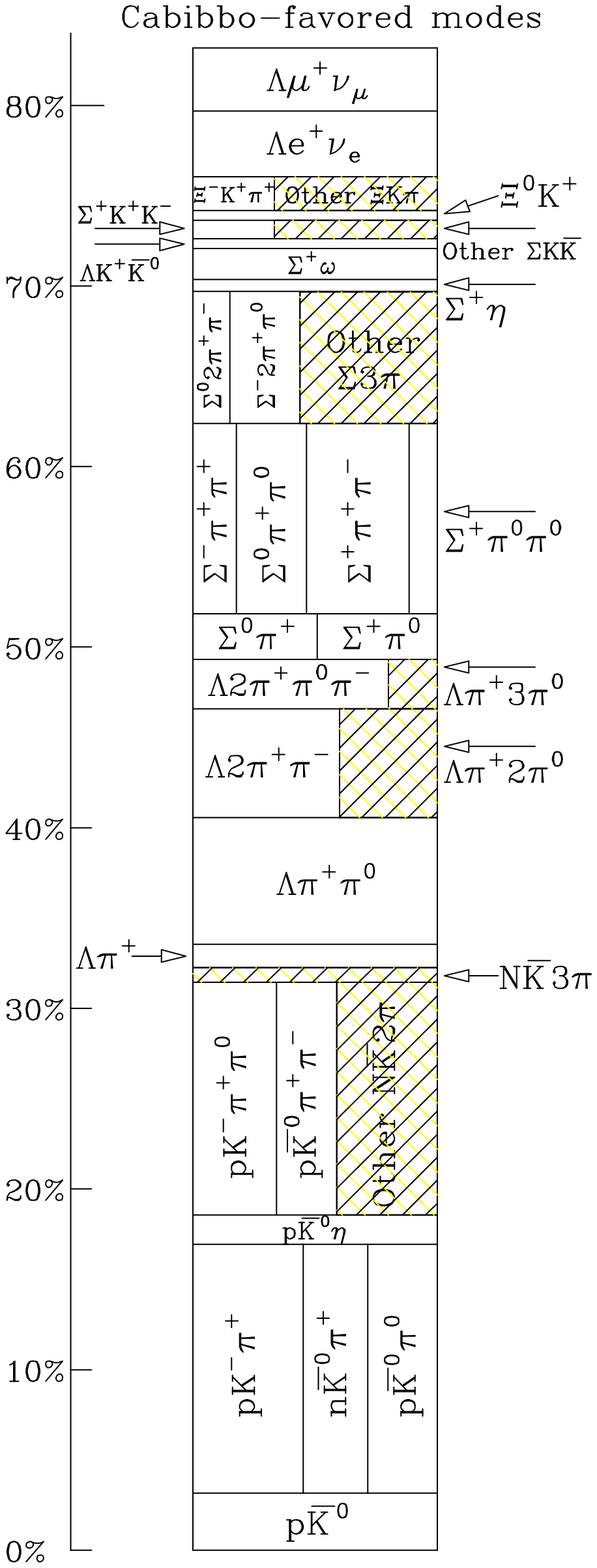}
\includegraphics[width=0.47\textwidth]{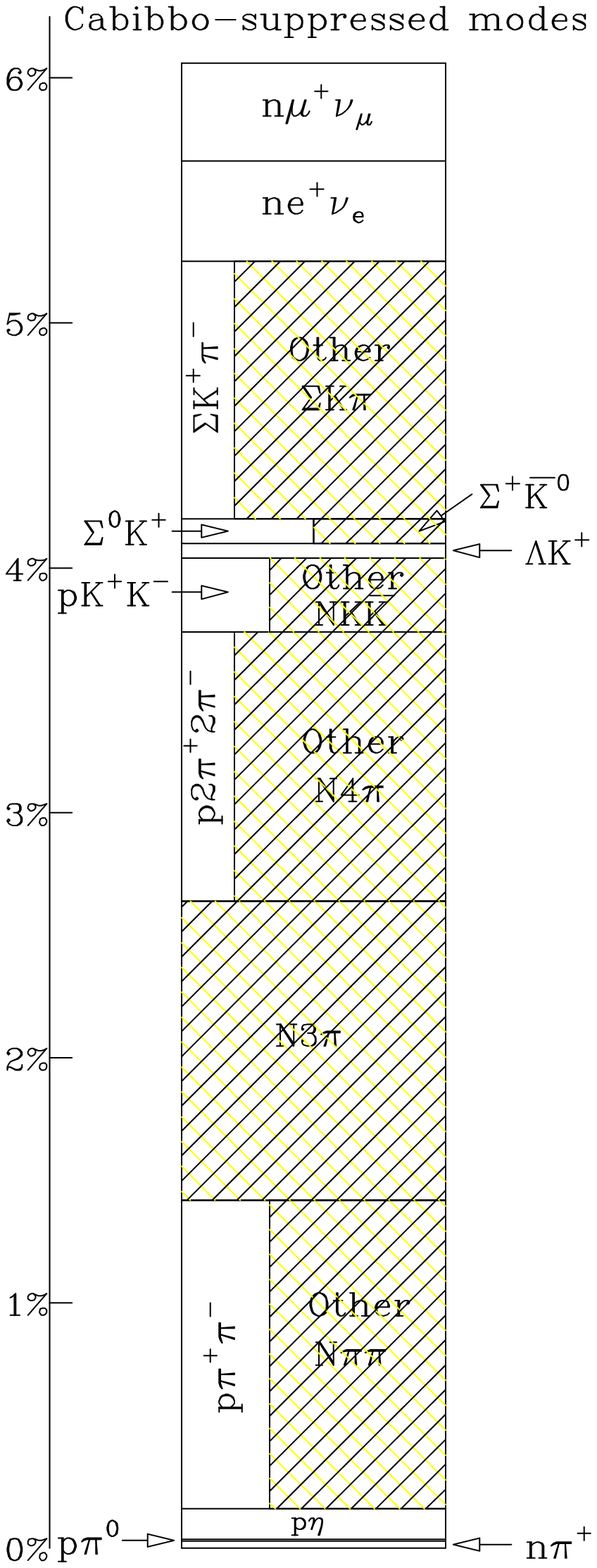}
\end{center}
\caption{Branching fractions of $\Lambda_c$ decays.  Left:  Cabibbo-favored
(CF, governed by weak transition $c \to s W^*$); right: Cabibbo-suppressed (CS,
giverned by weak transition $c \to d W^*$).
\label{fig:ldec}}
\end{figure}

$\Lambda_c$ branching fractions and their sources are listed in Tables
\ref{tab:cf} and \ref{tab:cs}.  These serve as inputs to Fig.\ \ref{fig:ldec},
in which the branching fractions are indicated by the areas of the boxes.
Shaded areas correspond to processes not represented by observed decays, but
whose rates are anticipated using a statistical isospin model \cite{GR18}.
The figures show only central values; errors are quoted in the tables.
\medskip

Some qualifying remarks are in order.  The $p K^- \pi^+$ decay mode, frequently
used to normalize others, is not firmly pinned down yet, with an $S$-value of
1.4 \cite{PDG18}.  The statistical isospin model is poorly obeyed for the
$N \bar K \pi$ and $\Sigma 3 \pi$ modes but well obeyed for the $\Sigma 2 \pi$
modes \cite{GR18}, possibly indicating the need to take account of resonant
substructure.  Nevertheless, one can draw some general conclusions.
\medskip

(1) We see a shortfall of about 10\% in accounting for all $\Lambda_c$ decays.
This could be filled in part by semileptonic decays to excited final states,
but a measurement ${\cal B}(\Lambda_c \to \Lambda e^+ \nu_e + X) = (3.95 \pm
0.34 \pm 0.09)\%$ by the BESIII Collaboration \cite{Ablikim:2018woi} limits
this possibility.
\medskip

(2) The Cabibbo-suppressed (CS) modes are not as well represented as the
Cabibbo-favored (CF) ones, though the anticipated totals are not far from the
expected ratio $|V_{cd}/V_{cs}|^2$, where $V_{ij}$ are elements of the
Cabibbo-Kobayashi-Maskawa matrix.
\medskip

(3) Modes involving neutrons, $\eta$, and $\eta'$ are under-represented.
\medskip

(4) There is sufficient phase space to accommodate higher-multiplicity modes,
such as $\Sigma 4\pi$ and $N 5\pi$, but no evidence for them has been
presented so far.
\medskip

(5) The statistical isospin model itself may be at fault.  Inclusive branching
fractions in $\Lambda_c$ decays would be very helpful in anticipating
as-yet-unseen modes without the help of models, as has been done for
$D_s$ decays \cite{PDGDs}.
\medskip

We urge more studies of $\Lambda_c$ decay modes containing neutrons, $\eta$,
and $\eta'$; greater investigation of the singly-Cabibbo-suppressed and
higher-multiplicity modes; and inclusive studies. Determination of resonant
substructure is a crucial ingredient in filling gaps only partially
addressed by an imperfect statistical isospin model. 
\medskip

\begin{table}
\caption{Branching fractions of CF $\Lambda_c$ decays.
\label{tab:cf}}
\begin{center}
\begin{tabular}{c c l} \hline \hline
Mode & Value (\%) & Source \\ \hline
$p \ok$ & $3.16 \pm 0.16$ & \cite{PDG18}$^a$ \\ \hline
$p K^- \pi^+$ & $6.23 \pm 0.33$ & \cite{PDG18} \\
$n \ok \pi^+$ & $3.64 \pm 0.50$ & \cite{PDG18}$^a$ \\
$p \ok \pi^0$ & $3.92 \pm 0.26$ & \cite{PDG18}$^a$ \\ \hline
$p \ok \eta$ & $1.6 \pm 0.4$ & \cite{PDG18}i$^a$ \\ \hline
$p K^- \pi^+ \pi^0$ & $4.42 \pm 0.31$ & \cite{PDG18} \\
$p \ok \pi^+ \pi^-$ & $3.18 \pm 0.24$ & \cite{PDG18}$^a$ \\
Other $N \overline{K} 2\pi$ & $5.28 \pm 0.39$ & \cite{GR18}$^{b,c}$ \\
\hline
$pK^- 2\pi^+ \pi^-$ & $0.14 \pm 0.09$ & \cite{PDG18} \\
Other $N \overline{K} 3\pi$ & $0.70 \pm 0.36$ & \cite{GR18}$^{b,c,d}$\\
\hline
$\Lambda \pi^+$ & $1.29 \pm 0.07$ & \cite{PDG18} \\
$\Lambda \pi^+ \pi^0$ & $7.0 \pm 0.4$ & \cite{PDG18} \\
$\Lambda 2\pi^+ \pi^-$ & $3.61 \pm 0.29$ & \cite{PDG18} \\
$\Lambda \pi^+ 2\pi^0$ & $2.41 \pm 0.13$ & \cite{GR18}$^{b,c}$ \\
$\Lambda 2\pi^+ \pi^0 \pi^-$ & $2.2 \pm 0.8$ & \cite{PDG18} \\
$\Lambda \pi^+ 3\pi^0$ & $0.55 \pm 0.2$ & \cite{GR18}$^{b,c,d}$ \\
\hline
$\Sigma^0 \pi^+$ & $1.28 \pm 0.07$ & \cite{PDG18} \\
$\Sigma^+ \pi^0$ & $1.24 \pm 0.10$ & \cite{PDG18} \\ \hline
$\Sigma^- \pi^+ \pi^+$ & $1.86 \pm 0.18$ & \cite{PDG18} \\
$\Sigma^0 \pi^+ \pi^0$ & $3.03 \pm 0.23$ & \cite{PDG18} \\
$\Sigma^+ \pi^+ \pi^-$ & $4.41 \pm 0.20$ & \cite{PDG18} \\
$\Sigma^+ \pi^0 \pi^0$ & $1.23 \pm 0.12$ & \cite{PDG18} \\ \hline
$\Sigma^0 2\pi^+ \pi^-$ & $1.10 \pm 0.30$ & \cite{PDG18} \\
$\Sigma^- 2\pi^+ \pi^0$ & $2.1 \pm 0.4$ & \cite{PDG18} \\
Other $\Sigma 3\pi$ & $4.1 \pm 0.5$ & \cite{GR18}$^{b,c,e}$ \\ \hline
$\Sigma^+ \eta$ & $0.69 \pm 0.23$ & \cite{PDG18} \\ \hline
$\Sigma^+ \omega$ & $1.69 \pm 0.21$ & \cite{PDG18} \\ \hline
$\Lambda K^+ \ok$ & $0.56 \pm 0.11$ & \cite{PDG18} \\ \hline
$\Sigma^+ K^+ K^-$ & $0.34 \pm 0.04$ & \cite{PDG18} \\
Other $\Sigma K \overline{K}$ & $0.68\pm0.34$ & \cite{GR18}$^{b,c}$ \\
\hline
$\Xi^0 K^+$ & $0.55 \pm 0.07$ & \cite{PDG18,Ablikim:2018bir}$^f$ \\
\hline
$\Xi^- K^+ \pi^+$ & $0.62 \pm 0.06$ & \cite{PDG18} \\
Other $\Xi K \pi$ & $1.24 \pm 0.12$ & \cite{GR18}$^{b,c,d}$ \\ \hline
$\Lambda e^+ \nu_e$ & $3.63 \pm 0.43$ & \cite{Ablikim:2015prg} \\
$\Lambda \mu^+ \nu_\mu$ & $3.49 \pm 0.53$ & \cite{Ablikim:2016vqd} \\ \hline
Total & $83.17 \pm 4.92$ \\
\hline \hline
\end{tabular}
\end{center}
\leftline{$^a$Branching fractions for modes with $\ok$ are obtained by
doubling those quoted for $K_S$.}
\leftline{$^b$Isospin statistical model \cite{GR18}. $^c$Subtraction of known
modes from estimated total.}
\leftline{$^d$Total estimated assuming equal branching fractions for
each charge state.}
\leftline{$^e\Sigma^+ \omega$ quoted separately. $^f$PDG value averaged
with new value from \cite{Ablikim:2018bir}.}
\end{table}

\begin{table}
\caption{Branching fractions of CS $\Lambda_c$ decays, in percent.
\label{tab:cs}}
\begin{center}
\begin{tabular}{c c c} \hline \hline
Mode & Value (\%) & Source \\ \hline
$p \pi^0$ & 0.008 & Theory \cite{Cheng:2018hwl} \\
$n \pi^+$ & 0.027 & Theory \cite{Cheng:2018hwl} \\ \hline
$p \eta$ & $0.124 \pm 0.030$ & \cite{PDG18} \\ \hline
$p \pi^+ \pi^-$ & $0.42 \pm 0.04$ & \cite{PDG18} \\
Other $N \pi \pi$ & $0.84 \pm 0.08$ & \cite{GR18}$^a$ \\ \hline
$N 3\pi$ & $1.22 \pm 0.30$ & \cite{GR18}$^b$ \\ \hline
$p 2\pi^+ 2\pi^-$ & $0.22 \pm 0.14$ & \cite{PDG18} \\
Other $N 4\pi$ & $0.88 \pm 0.56$ & \cite{GR18}$^a$ \\ \hline
$p K^+ K^-$ & $0.10 \pm 0.04$ & \cite{PDG18} \\
Other $N K^+ K^-$ & $0.20 \pm 0.08$ & \cite{GR18}$^a$ \\ \hline
$\Lambda K^+$ & $0.06 \pm 0.012$ & \cite{PDG18} \\ \hline
$\Sigma^0 K^+$ & $0.051 \pm 0.008$ & \cite{PDG18} \\
$\Sigma^+ K^0$ & $0.051 \pm 0.008$ & \cite{GR18}$^a$ \\ \hline
$\Sigma^+ K^+ \pi^-$ & $0.21 \pm 0.06$ & \cite{PDG18} \\
Other $\Sigma K \pi$ & $0.84 \pm 0.24$ & \cite{GR18}$^a$ \\ \hline
$n e^+ \nu_e$ & $0.41 \pm 0.03$ & Lattice QCD \cite{Meinel:2017ggx} \\
$n \mu^+ \nu_\mu$ & $0.40 \pm 0.03$ & Lattice QCD \cite{Meinel:2017ggx} \\
\hline
Total & $6.06 \pm 0.84$ & \\
\hline \hline
\end{tabular}
\end{center}
\leftline{$^a$ Total estimated assuming equal branching fractions for each
charge state.}
\leftline{$^b$Branching ratio to $p \pi^+ \pi^0 \pi^-$ taken as $(0.304 \pm
0.076)\%$ (geometric mean of $p \pi^+ \pi^-$}
\leftline{and $p 2\pi^+ 2\pi^-$ modes) multiplied by 4 for total number of
charge states.}
\end{table}

We thank Roy Briere, Alexander Gilman, Hai-Bo Li, Xiao-Ryu Lyu, Stefan Meinel,
Hajime Muramatsu, and Ruth Van de Water for useful communications.  M.G. thanks
the CERN TH division and J.L.R. thanks the Technion for kind hospitality.

\end{document}